\documentstyle[aps,prb,eqsecnum,epsf,preprint,tighten]{revtex} 
\begin{document}
\preprint{NSF-ITP-98-019}
\title{Boundary Critical Phenomena in the Three-State Potts Model}
\author{Ian Affleck$^{1}$, Masaki Oshikawa$^2$ and Hubert 
Saleur$^3$}
\address{$^1$Canadian Institute for Advanced
Research  and Department of Physics and Astronomy, University of 
British 
Columbia, 
Vancouver, B.C., Canada, V6T 1Z1 \\ 
$^2$Department of Physics, Tokyo Institute of Technology, \\ Oh-
okayama, 
Meguro-ku, 
 Tokyo 152-8551, Japan\\ 
$^3$ Department of Physics, University of Southern California, Los 
Angeles, 
CA90089-
0484, USA}
\maketitle
\begin{abstract}Boundary critical phenomena are studied in the 3-
State Potts
model in 2 dimensions using conformal field theory, duality and
renormalization group methods. A presumably complete set of 
boundary
conditions is obtained using both fusion and orbifold methods.  
Besides the
previously known free, fixed and mixed boundary conditions a new 
one
is obtained.  This illustrates the necessity of
considering fusion with operators that don't occur in the bulk
spectrum, to obtain all boundary conditions. It is shown that this new 
boundary 
condition 
is dual to the mixed ones.  The phase diagram for the
quantum chain version of the Potts model is analyzed using duality 
and 
renormalization 
group arguments.
\end{abstract}
\section{Introduction}
Recently there has been considerable interest in the behaviour of two 
dimensional 
systems with boundaries, in the context of string theory, classical 
statistical 
mechanics 
and quantum impurity problems.  Exact results on the critical 
behavior of these 
systems 
have been obtained using boundary conformal field 
theory (CFT).\cite{Cardy,Affleck}  
More 
complete 
exact results on universal crossover functions have also been obtained 
using exact 
S-
matrix methods.\cite{Zamolodchikov1}  One of the simplest 
examples of such a 
system is provided by 
the 
3-
state Potts model.  
It can be related, via conformal embeddings,\cite{Affleck2} to 
quantum Brownian 
motion on a hexagonal lattice\cite{Yi} and to tunnelling in quantum 
wires.\cite{Kane}  
The classical Hamiltonian for this model can be written by 
introducing an angular 
variable at each site of a square lattice, $\theta_{i}$, restricted to take 
only 3 
values: 
$0,\pm 2\pi /3$.
\begin{equation} \beta H = -J\sum_{<i,j>}\cos (\theta_i-\theta_j).
\label{class} \end{equation}

When the model is at its critical coupling, $J_c$, various universality 
classes of 
boundary critical phenomena are possible.  These  include  free 
boundary 
conditions 
(b.c.'s) and (3 different) fixed b.c.'s, $\theta_i=0$ (or $2\pi /3$ or $-
2\pi /3$), for 
$i$ on 
the boundary.  In addition, it was argued\cite{Saleur1} that there are 
also 3 
``mixed'' b.c.'s in 
which 
one of the 3 spin states is forbidden at the boundary so that the Potts 
spins on the 
boundary fluctuate between two of the states (for example, between 
$2\pi /3$ and 
$-2\pi 
/3$). 

In what follows it will be convenient to also consider the standard 
quantum chain 
representation.  The Hamiltonian is written in terms of unitary 
matrices,
$M_i$ and $R_i$ defined at each site.  
\begin{equation} 
M=\pmatrix{0&1&0 \cr 0&0&1 \cr 1&0&0 \cr},\ \  
R=\pmatrix{e^{2\pi 
i/3}&0&0\cr 
0&e^{4\pi i/3}&0
\cr 0&0&1 \cr}
\end{equation}

In fact, these 2 matrices can be transformed into each other:
\begin{equation} R = U^\dagger M U.\end{equation}
This is related to the duality symmetry.  The Hamiltonian is
\begin{equation}
H=-\sum_i[(M_i+M_i^\dagger )+(R_i^\dagger
R_{i-1}+R_{i-1}^\dagger R_i)].
\label{Ham}\end{equation}
Note that the second term corresponds to the classical Potts model 
with the 3 
different 
states corresponding to the vectors (1,0,0), (0,1,0) and (0,0,1). The 
first term flips 
the 
spin on each site between the 3 states.  It is like a transverse field in 
the Ising 
model. 
The model has $Z_3$ symmetry which interchanges the 3 basis 
vectors. 
Decreasing 
the 
strength of the transverse field term puts the system in the ordered 
phase; 
increasing it 
gives the disordered phase.  As written, these terms exactly balance; 
the model is 
at 
its 
critical point.  One way of seeing this is to observe that, for this value 
of the 
coupling 
constant, the Hamiltonian maps into itself under the duality 
transformation:
\begin{eqnarray}
R'_{i+1/2}&\equiv& \prod_{j=0}^iM_j\nonumber \\
M'_{i+1/2}&\equiv& R_{i+1}^\dagger 
R_i.\label{dual}\end{eqnarray}

The 6 fixed and mixed b.c.'s were represented in terms of boundary
states.\cite{Cardy}  
These are defined by a modular transformation of the partition 
function on a 
cylinder of 
circumference $\beta$ and length $l$ with b.c.'s A and B  
at the two 
ends:
\begin{equation} 
Z_{AB}=\hbox{tr}\exp [-\beta H^l_{AB}]=<A |\exp [-lH^\beta_P]|B 
>.
\end{equation}
Here, $H^l_{AB}$ is the Hamiltonian on a strip of length $l$ with 
b.c.'s $A$ and 
$B$ at 
the 2 ends.  $H^\beta_P$ is the Hamiltonian on a circle of 
circumference $\beta$. 
$Z_{AB}$ may be expanded in characters of the (chiral) Virasoro 
algebra:
\begin{equation}
Z_{AB}=\sum_kn^k_{AB}\chi_k(q).\end{equation}
Here $q$ is the modular parameter, $q\equiv \exp [-\pi \beta /l]$, $k$ 
labels 
(chiral) 
conformal towers, $\chi_k$ are the characters and $n^k_{AB}$ are 
non-negative 
integers.  The boundary states may be expanded in Ishibashi states, 
constructed 
out 
of 
each conformal tower:
\begin{equation} |A>=\sum_k|k><k,0|A >.\end{equation}
One way of generating a complete set of boundary states (and hence 
b.c.'s) from 
an 
appropriately chosen reference state is by fusion.  Beginning with the 
reference 
boundary 
state $|\tilde 0>$,  one constructs a set of boundary states, $|\tilde j>$ 
associated 
with the  
conformal towers, $j$.  Its matrix elements are given by:
\begin{equation}
<i,0|\tilde j>={S^i_j\over S^i_0}<i,0|\tilde 0>,\label{fusion}\end{equation}
where $S^i_j$ is the modular S-matrix.  
This construction gives physically sensible multiplicities, 
$n_{AB}^i$; that is 
they  
are 
non-negative integers obeying $n_{AA}^0=1$.  This construction 
relies on the 
Verlinde 
formula\cite{Verlinde} which relates the modular S-matrix to the 
fusion rule 
coefficients.

A subtlety arises in the Potts model connected with an extended W-
algebra.  
While 
there 
are 10 Virasoro conformal towers for central charge c=4/5, labeled by 
pairs of 
integers, 
$(n,m)$ with $1\leq n \leq 4$ and $m\leq n$, only 4 larger conformal 
towers, 
which 
are 
combinations of these ones, occur in the bulk spectrum or with certain 
pairs of 
b.c.'s.  
Furthermore, two of these conformal towers occur twice in the bulk 
spectrum 
corresponding to pairs of operators of opposite charge ($\pm 1$) with 
respect to 
the 
$Z_3$ symmetry of the Potts model.  [In general, operators can have 
charge q=0, 
1 
or -1, 
transforming under $Z_3$ transformations as:
\begin{equation} {\cal O}\to e^{iq\theta}{\cal O},\end{equation}
for $\theta = 0$, $\pm 2\pi /3$.]  These operators are $\sigma$, 
$\sigma^\dagger$ of 
dimension 1/15 and $\psi$, $\psi^\dagger$ of dimension 2/3.  The 
Potts model 
also 
contains an energy operator, $\epsilon$ of dimension 2/5 as well as 
the identity 
operator, 
$I$.  These W-characters  are given by:
\begin{eqnarray}
\chi_I&=&\chi_{11}+\chi_{41},\ \ \chi_\epsilon = 
\chi_{21}+\chi_{31}
\nonumber \\
\chi_\sigma&=&\chi_{\sigma^\dagger}=\chi_{33},\ \ \chi_\psi = 
\chi_{\psi^\dagger}=\chi_{43},\end{eqnarray}
where $\chi_{nm}$ is the Virasoro character for the (n,m) conformal 
tower.
The Potts model has a fusion algebra which closes on these operators.  
The 
modular 
transform of these W-characters can be expressed entirely in terms of 
W-
characters 
and 
the corresponding S-matrix and fusion rule coefficients obey the 
Verlinde 
formula.  
Ambiguities in the S-matrix and fusion rules associated with having 
operators of 
equal 
dimension are removed by requiring consistency with the $Z_3$ 
symmetry.
Cardy constructed a set of boundary states which were linear 
combinations of the 
Ishibashi states constructed using the extended W-algebra.  The 
reference state for 
the 
fusion process, in this construction, is the boundary state, $|\tilde I>$, 
obeying 
$Z_{\tilde I\tilde I}=\chi_I$.  It was argued in [\onlinecite{Cardy}] 
that it
corresponds to one of the fixed b.c.'s, the other two being $|\tilde \psi 
>$ and 
$|\tilde \psi^\dagger >$.  Similarly $|\tilde \epsilon >$, $|\tilde \sigma 
>$ and 
$|\tilde 
\sigma^\dagger >$ correspond to the 3 mixed b.c.'s.  All partition 
functions 
involving 
these 6 b.c.'s can be expressed in terms of W characters.  On the other 
hand, it was 
observed that partition functions that combine free b.c.'s with fixed or 
mixed 
cannot be 
expressed in terms of W characters and the corresponding free 
boundary state was 
not 
determined.

Clearly the set of b.c.'s generated by fusion with the primary fields of 
the bulk 
Potts 
spectrum (which are covariant with respect to the W-algebra) is not 
complete, 
since 
it 
doesn't include the free b.c..  In the next section we will generate a 
presumably
complete set of boundary states including the one corresponding to 
free
boundary conditions and one new boundary state. We do this two 
different
ways; one method uses fusion and the other uses an orbifold 
projection.
In the final section we will
explore the physical significance of this new boundary condition and 
the
boundary renormalization group flow diagram.  The appendix 
contains a 
peripherally
related result: a general proof that  the groundstate entropy always 
{\it increases} under
fusion.

\section{Boundary States}\subsection{Fusion Approach}
In order to determine the ``free'' boundary state and 
check 
for possible additional boundary states (and conditions) we must work 
with the 
larger set 
of conformal towers not constrained by the W-symmetry.  The full 
modular S-
matrix, in 
the space of all 10 Virasoro conformal towers
 that can occur in a c=4/5 minimal model, is given in Table 1.  
The state 
$|\tilde 
I>$ 
may be expanded in terms of W-Ishibashi states as:
\begin{equation} |\tilde I>=N\{ |I>+|\psi >+|\psi^\dagger >+\lambda 
[|\epsilon 
>+|\sigma 
>+|\sigma^{\dagger} >]\} ,\end{equation}
where 
\begin{equation} N^4={5-\sqrt{5}\over 30}, \ \ 
\lambda^2={1+\sqrt{5}\over 
2}.\end{equation}
The W-Ishibashi states may be expanded in terms of  Virasoro 
Ishibashi states as:
\begin{equation} |I>=|11>+|41>,\ \ |\epsilon 
>=|21>+|31>.\end{equation}

Now consider all new boundary states that can be obtained from 
$|\tilde I>$ by 
fusion 
with all 9 non-trivial Virasoro primaries using Eq. (\ref{fusion}).
 Note that $|\tilde I>$ has 
zero amplitude 
for the 
last 4 Ishibashi states in Table I: (4,4), (4,2), (2,2), (3,2).  Also note 
that the 
$S^{(1,1)}_i=S^{(4,1)}_i$ for all i except for these last 4 states.  The 
same 
statement 
holds for $S^{(2,1)}_i$ and $S^{(3,1)}_i$.  Thus expanding the 
identity tower 
with 
respect to the W-algebra into (1,1) and (4,1) doesn't lead to any 
additional 
boundary 
states.  Neither does expanding the $\epsilon$ tower into (2,1) and 
(4,3).  The 
(4,3) 
and 
(3,3) towers just give the states found previously since these are 
themselves W-
towers.  
However, two additional boundary states can be obtained by fusion 
with (4,4) and 
(2,2).  
On the other hand, fusion with (4,2) gives the same result as (4,4) and 
(3,2) the 
same as 
(2,2) since all but the last 4 elements in the corresponding rows
in Table I are equal.  
Thus, the fusion construction, beginning with the W-invariant 
boundary state 
$|\tilde I>$ 
but considering the full set of Virasoro primaries leads to 2
additional boundary states 
besides the 6 found previously by considering fusion with W 
primaries. 

Note that we have performed a sort of hybrid construction.  We could 
have
instead begun 
with the reference boundary state $|\tilde {11}>$ such that
$Z_{\tilde {11},\tilde {11}}=\chi_{11}$.  In this case we would  
obtain a larger set 
of boundary states.  
However, these states do not occur in the Potts model.
One reason is that $|\tilde {11}>$ 
is not consistent with $|\tilde I>$.  This follows from the identity:
\begin{equation} |\tilde I>={1\over \sqrt{2}}[|\tilde {11}>+|\tilde 
{41}>].\label{I}\end{equation}
The factor of $1/\sqrt{2}$ in Eq. (\ref{I}), necessary to avoid a 2-fold 
degeneracy 
in the 
spectrum of $Z_{\tilde I \tilde I}$, leads to an unphysical partition 
function 
$Z_{\tilde 
I\tilde {11}}$, with non-integer multiplicities.  Another reason why 
this larger set 
of 
boundary states cannot occur in the Potts model is because they 
contain Ishibashi 
states 
not derived from the bulk spectrum.
The 8 boundary states discussed here presumably form a complete set 
of states 
which are 
mutually consistent.

We note that the idea of obtaining new boundary states (and 
conditions)
by fusion with
operators which don't occur in the bulk spectrum is also fundamental 
to
the
solution of the non-Fermi liquid fixed points in the Kondo
problem.\cite{Affleck3}  In that case, the reference  state
was chosen to give a Fermi liquid b.c..  The conformal embedding
representing
the free fermions restricts the bulk spectrum to contain only certain
products of operators from
the spin, charge and flavour sectors.  Fusion with pure spin operators,
not
contained in the bulk spectrum, gives the infrared stable fixed points
of both Fermi liquid and non- Fermi liquid variety.

The two additional boundary states for the Potts model, found above,
are:
\begin{eqnarray}
|\tilde{44}>&=&N\sqrt{3}[(|11>-|41>)-\lambda (|21>-|31>)]
\nonumber \\
|\tilde{22}>&=&N\sqrt{3}[\lambda^2 (|11>-|41>)+\lambda^{-1}
(|21>-|31>)].
\end{eqnarray}
The partition functions for any pair of b.c.'s can be determined from
the boundary
states
using:
\begin{equation} <i|\exp {-lH^\beta_P}|j>=\delta_{ij}\chi_i(\tilde
q),\end{equation}
where $\tilde q=e^{-4\pi l/\beta }$.  Finally we perform a modular
transformation
to the
q-representation:
\begin{equation} \chi_i(\tilde q)=\sum_jS^j_i\chi_j(q).\end{equation}
Alternatively, we may determine these partition functions from the
fusion rule
coefficients. For a b.c., $\tilde i$ obtained by fusion with primary
operator, ${\cal
O}_i$
from $|\tilde I>$ and some other b.c., $\tilde j$,
\begin{equation}
n^k_{\tilde i \tilde j}=\sum_lN^k_{il}n^l_{\tilde I \tilde
j}.\end{equation}
Here $N^k_{il}$ is the number of times that the primary operator
${\cal O}_k$
occurs
in the operator product expansion of ${\cal O}_i$ with ${\cal O}_l$.
The needed
fusion
rule coefficients are given in Table 2. These are derived from the
fusion rules of
the
tetracritical Ising model.  For instance, to get the first box in the
table
we use:
\begin{equation} {\cal O}_{44}\cdot I={\cal O}_{44}\cdot [{\cal
O}_{11}+{\cal
O}_{41}]\to {\cal O}_{44}+{\cal O}_{42}.\end{equation}  In cases
where two
dimension 2/3 (1/15) operators occur in the O.P.E. we have
interpreted them as
$\psi+\psi^\dagger$ ($\sigma + \sigma^{\dagger}$).
This calculation shows that all partition functions involving
$|\tilde{44}>$ and
any
of
the   fixed or mixed boundary states are the same as those determined
previously
for the
free b.c.. Hence we conclude that $|\tilde{44}>$ is the free boundary
state.  On
the
other
hand, $|\tilde {22}>$ is a new boundary state corresponding to a new
b.c. whose
physical
interpretation is so far unclear.  In the next section we investigate
the
nature
of this new boundary fixed point.  First, however, we obtain this set
of boundary states by an interesting different method.
\subsection{Orbifold Approach}
An alternative way of producing the complete set of boundary states
for the Potts model is based on obtaining the Potts model from an
orbifold projection on the other c=4/5 conformal field theory, the
tetracritical Ising model, which has a diagonal bulk partition
function.\cite{Difrancesco}
A $Z_2$ Ising charge, $q_i$, can be assigned to each primary
operator, ${\cal O}_i$ of the tetracritical Ising model which is 0 for
the first 6 entries in Table I and 1 for the remaining 4.  [This is a 
special case of a general construction for minimal models.  Choosing
a different fundamental domain for Kac labels, (n,m) with
\begin{equation}
1\leq n\leq p'-1,\ \  1\leq m \leq p-1,\ \ n+m=0\ \  \hbox{mod}\ 2,
\end{equation} the charge is:
\begin{equation}
q=n+1.\end{equation}
The c=4/5 case corresponds to $p=6$, $p'=5$. This identification is
consistent with the Landau-Ginsburg description of the tetracritical
Ising model.\cite{Zamolodchikov2}  The charge 1
operators ${\cal O}_{22}$,
${\cal O}_{44}$ and ${\cal O}_{32}$ correspond to $\phi$, 
$:\phi^3:$ and
$:\phi^5:$ respectively.  The charge 0 operators ${\cal O}_{33}$,
${\cal O}_{21}$ and ${\cal O}_{43}$ correspond to $:\phi^2:$, 
$:\phi^4:$ and $:\phi^6:$ respectively. The other 3 operators other than the
identity presumably could be identified with operators in the Landau
Ginsburg description containing derivatives with the number of 
powers of
$\phi$ even for ${\cal O}_{41}$ and ${\cal O}_{31}$ and odd for 
${\cal
O}_{42}$.]
The tetracritical Ising model has the diagonal bulk partition function:
\begin{equation} Z_{TC}\equiv 
Z_{++}=\sum_{i=1}^{10}|\chi_i|^2,\end{equation}
where we number the conformal towers from 1 to 10 in the order in 
Table I.
We may define a twisted partition function:
\begin{equation}
Z_{+-}\equiv \sum_{i=1}^{10}(-1)^{q_i}|\chi_i|^2.\end{equation}
We also define 2 other twisted partition functions by the modular 
transforms
of $Z_{+-}$:
\begin{equation} Z_{-+}\equiv {\cal S}Z_{+-},\ \  Z_{--}={\cal 
T}Z_{-
+},\end{equation}
where ${\cal S}$ is the modular transformation $\tau \to -1/\tau$
and ${\cal T}$ is the modular transformation $\tau \to \tau +1$.
It can be shown that:
\begin{equation}
Z_{Potts}=Z_{orb}=(1/2)[Z_{++}+Z_{+-}+Z_{-+}+Z_{--
}].\end{equation}
We may think of the first two terms as representing the contribution 
of the 
untwisted
sector of the Hilbert Space, with the $Z_2$ invariant states projected 
out.
The second two terms represent the contribution of the twisted sector 
of
the Hilbert Space, corresponding
to twisted boundary conditions on the circle.  [For the simpler case of 
the
c=1 bosonic orbifold the twisted boundary conditions are simply $\phi 
(0)=-\phi 
(l)$.]
These contributions are explicitly:
\begin{eqnarray}
(1/2)[Z_{++}+Z_{+-
}]&=&|\chi_{11}|^2+|\chi_{41}|^2+|\chi_{21}|^2+|\chi_{31}|^2+|\chi
_{43}|^2+|\
\chi_{33}|^2\nonumber \\
(1/2)[Z_{-+}+Z_{--}]&=& 
\bar\chi_{11}\chi_{41}+\bar\chi_{41}\chi_{11}+\bar\chi_{21}\chi_{
31}
+\bar\chi_{31}\chi_{21}+|\chi_{43}|^2+|\chi_{33}|^2.
\end{eqnarray}

There are two types of Ishibashi states which may be used to construct 
boundary 
states in
the orbifold model.  We may take states from the untwisted sector,
projecting out the $Z_2$
invariant parts or we may take states from the twisted sector. 
The first set of Ishibashi
states are labeled by the first 6 ($Z_2$ even) conformal towers in 
Table I.
We refer to
these untwisted $|43>$ and $|33>$ states as $|\psi_u>$ and 
$|\sigma_u>$ 
respectively.  There
are 2 additional Ishibashi states from the twisted sector, $|\psi_t>$ 
and 
$|\sigma_t>$.
We then define:
\begin{eqnarray} |\psi >&\equiv& 
(1/\sqrt{2})[|\psi_u>+i|\psi_t>]\nonumber \\
|\psi^\dagger>&\equiv& (1/\sqrt{2})[|\psi_u>-i|\psi_t>]\end{eqnarray}
and similarly for $|\sigma>$.
One way of constructing consistent boundary states, using only the 
untwisted 
sector, is by projecting out the $Z_2$ even parts of the tetracritical 
Ising 
boundary states.  From inspecting Table I we see that the various
tetracritical Ising boundary states are mapped into each other by the
$Z_2$ transformation.  We have ordered them in Table I so that
successive pairs are interchanged, apart from $|\tilde {43}>$ and 
$|\tilde {33}>$
which are invariant. We expect the conjugate pairs to correspond to
various generalized spin-up and spin-down boundary conditions.
We may formally write the transformed states as:
\begin{equation} (-1)^{\hat Q}|A_{TC}>.\end{equation}
With each
boundary state, $|A_{TC}>$, of the tetracritical Ising model, we may
associate a boundary state, $|A_{Potts}>$ of the Potts model using:
\begin{equation}
<i,0|A_{Potts}>={1+(-1)^{q_i}\over \sqrt{2}}<i,0|A_{TC}>.
\end{equation}
Formally we may write:
\begin{equation} |A_{Potts}>={1+(-1)^{\hat Q}\over 
\sqrt{2}}|A_{TC}>.\end{equation}
It is necessary to divide by $\sqrt{2}$ in order that the identity
operator only appear once in the diagonal partition functions.
In this way we obtain the following Potts boundary states from each
tetracritical Ising boundary state:
\begin{eqnarray}
|\tilde {11}_{TC}>&\to&|\tilde I>\nonumber \\
|\tilde {41}_{TC}>&\to& |\tilde I>\nonumber \\
|\tilde {31}_{TC}>&\to&|\tilde \epsilon >\nonumber \\
|\tilde {31}_{TC}>&\to&|\tilde \epsilon >\nonumber \\
|\tilde {43}_{TC}>&\to&|\tilde \psi>+|\tilde
\psi^\dagger>\nonumber \\
|\tilde {33}_{TC}>&\to&|\tilde \sigma >+|\tilde
\sigma^\dagger >\nonumber \\ 
|\tilde {44}_{TC}>&\to& |\tilde {44}>\nonumber \\
|\tilde {42}_{TC}>&\to&|\tilde {44}>\nonumber \\
|\tilde {22}_{TC}>&\to&|\tilde {22}>\nonumber \\
|\tilde {32}_{TC}>&\to&|\tilde {22}>
\end{eqnarray}
[Note that the states $|\tilde {11}_{TC}>$ and $|\tilde {41}_{TC}>$
are the same states simply labeled $|\tilde {11}>$ and $|\tilde {41}>$
in Eq. (\ref{I}).]
We observe that this construction gives us a sum of Potts boundary 
states
in the 43 and 33 cases
 because the corresponding tetracritical Ising
boundary states are $Z_2$ invariant.
We may remedy this situation by forming linear combinations of the 
projected 
tetracritical
boundary states with the twisted Ishibashi states:
\begin{eqnarray}|\tilde \psi >&=&(1/2){1+(-1)^{\hat Q}\over 
\sqrt{2}}|\tilde 
{43}_{TC}>-N\sqrt{3/2}
[|\psi_t>+\lambda|\sigma_t>]\nonumber \\
|\tilde \psi^\dagger >&=&(1/2){1+(-1)^{\hat Q}\over \sqrt{2}}|\tilde 
{43}_{TC}>+N\sqrt{3/2}
[|\psi_t>+\lambda|\sigma_t>]\nonumber \\
|\tilde \sigma >&=&(1/2){1+(-1)^{\hat Q}\over \sqrt{2}}|\tilde 
{33}_{TC}>-
N\sqrt{3/2}
[\lambda^2|\psi_t>-\lambda^{-1}|\sigma_t>]\nonumber \\
|\tilde \sigma^\dagger >&=&(1/2){1+(-1)^{\hat Q}\over 
\sqrt{2}}|\tilde 
{33}_{TC}>+N\sqrt{3/2}
[\lambda^2|\psi_t>-\lambda^{-1}|\sigma_t>].
\end{eqnarray}
This construction is rather reminiscent of the one used to obtain 
orbifold 
boundary states
to describe a defect line in the Ising model\cite{Oshikawa} where it
was
also necessary to add a contribution from the twisted sector when the
periodic boson boundary states were invariant under the $Z_2$
transformation.  This is presumably an important ingredient of a 
general
prescription for constructing boundary states for orbifold models.
\section{The New Boundary Condition}
The various partition functions involving the new b.c. are given 
below.  
Henceforth, to 
simplify our notation, we will refer to the fixed b.c.'s as A,B and C 
(corresponding 
to the 
three possible states of the Potts variable) the mixed b.c.'s as AB, AC, 
BC, the free 
b.c. 
as ``free'' and the new b.c. corresponding to the $|\tilde {22}>$ 
boundary state as 
``new''.  
(In [\onlinecite{Cardy}] the notation ``A+B'' was used rather than 
``AB''.)
\begin{eqnarray} Z_{new,A}&=& Z_{new,B}= 
Z_{new,C}=\chi_{22}+\chi_{32}=Z_{free,AB}\nonumber \\
Z_{new,AB}&=& Z_{new,BC}= 
Z_{new,AC}=\chi_{44}+\chi_{42}+\chi_{22}+\chi_{32} \nonumber 
\\
Z_{new,free}&=&\chi_{\epsilon}+\chi_{\sigma}+\chi_{\sigma^\dagger}
=Z_{AB,A}+Z_{AB,B}+Z_{AB,C}\nonumber \\
Z_{new,new}&=&\chi_I+\chi_\epsilon +\chi_\sigma 
+\chi_{\sigma^\dagger}+\chi_{\psi}+\chi_{\psi^\dagger}
=Z_{AB,AB}+Z_{AB,BC}+Z_{AB,AC}.\label{PFs}\end{eqnarray}
Several clues to the nature of the new fixed point are provided by 
these partition 
functions.  The equality of the three partition functions on the first 
line of Eq. 
(\ref{PFs}) 
and on the second line strongly suggests that the new b.c. is $Z_3$ 
invariant.  This 
is 
also probably implied by the fact that their is only 1 new b.c., not 3.  
In general, 
the 
diagonal partition functions, $Z_{\alpha \alpha}$ give the boundary operator 
content with 
b.c. 
$\alpha$, 
with the usual relation between the finite-size energies and the scaling 
dimensions 
of 
operators. This in turn gives information about the renormalization 
group stability 
of the 
boundary fixed point.  We give all diagonal partition functions below:
\begin{eqnarray}
Z_{A,A}&=&\chi_I \nonumber \\
Z_{AB,AB}&=&\chi_I+\chi_\epsilon \nonumber \\
Z_{free,free}&=&\chi_I+\chi_\psi + \chi_{\psi^\dagger} \nonumber \\
Z_{new,new}&=&\chi_I+\chi_\epsilon +\chi_\sigma 
+\chi_{\sigma^\dagger}+\chi_{\psi}+\chi_{\psi^\dagger}.\label{PFD}
\end{eqnarray}
We see that the fixed boundary fixed point is completely stable.  
Apart from the 
identity 
operator it only contains operators of dimensions $\geq 2$.  The 
mixed fixed point 
has 1 
relevant operator of dimension 2/5 while the free fixed point has 2 
relevant 
operators, 
both of dimension 2/3.  It is easy to see, on physical grounds, what 
these operators 
are.  
Consider adding a boundary ``magnetic field'' to the free b.c.:
\begin{equation} \beta H\to \beta H - \sum_j'[h e^{i\theta_j}+ 
c.c.].\end{equation}
Here the sum runs over the spins on the boundary only.  $h$ is a 
complex field 
and 
$c.c.$ denotes complex conjugate. The two relevant operators at the 
free fixed 
point 
correspond to the real and imaginary parts of $h$.  If we assume that 
$|h|$ 
renormalizes to 
$\infty$ then it would enforce a fixed b.c. for generic values of arg 
($h$).  For 
instance, a 
real positive $h$ picks out $\theta_j=0$.  There are three special 
directions, 
arg ($h$)=$\pi$, 
$\pm \pi /3$ for which two of the Potts states remain degenerate.  For 
instance, for 
$h$ real 
and negative, $\theta = \pm 2\pi /3$. 
  These values of Im ($h$) are
invariant under renormalization due to a $Z_2$ symmetry.  We expect
the system to renormalize to the mixed fixed point for these values
of arg ($h$).
  Im ($h$) corresponds to the single relevant 
coupling constant at 
the 
mixed fixed point with Im ($h$)=0. Giving $h$ a small imaginary part at
this fixed point will select one of the 2 Potts states $2\pi /3$ or
$-2\pi /3$, corresponding to an RG flow from mixed to fixed.
 Since the free fixed point has $Z_3$ symmetry we 
can classify 
the 
relevant operators by their $Z_3$ charge. 
The two operators at the free fixed point, $e^{\pm i\theta_j}$, have 
charge $\pm 
1$, 
corresponding to $\psi$ and $\psi^\dagger$.  

We see from Eq. (\ref{PFD}) that there are 5 relevant operators at the 
new fixed 
point. 
Two with charge 1, two with charge -1 and 1 with charge 0.  The 
charged 
operators 
presumably arise from applying a magnetic field.  However, even if 
we preserve 
the 
$Z_3$ symmetry, there still remains 1 relevant operator, $\epsilon$ of 
dimension 
2/5.  
Thus, we might expect the new fixed point to be unstable, even in the 
presence of 
$Z_3$ 
symmetry, with an RG flow to the free fixed point occurring.  

It turns out that there is a simple physical picture of the new boundary 
condition 
within the
quantum Potts chain realization.  The corresponding classical model 
can also be 
constructed but involves
negative Boltzmann weights.  Therefore we first discuss the quantum 
model and 
turn to the
classical model at the end.

We now consider the 
quantum 
chain model on a finite interval, $0\leq i \leq l$.  In order to explore 
the $Z_3$ 
symmetric part of the phase diagram it is convenient to consider the 
model with a 
complex transverse field, $h_T$ at the origin and a free b.c. at $l$:
\begin{equation}
H=-(h_TM_0+ h_T^*M_0^\dagger)-\sum_{i=1}^l[(M_i+M_i^\dagger 
)+(R_i^\dagger 
R_{i-1}+R_{i-1}^\dagger R_i)].
\end{equation}
We can effectively map out the phase diagram by considering the 
duality  
transformation 
of Eq. (\ref{dual}).  The dual lattice consists of the points $i+1/2$ for 
$i=0,1,\ldots 
l$.
Note that, from Eq. (\ref{dual}):
\begin{equation} R'_{1/2}\equiv M_0.\end{equation}
The exactly transformed Hamiltonian is:
\begin{equation}
H= -(h_TR{'}_{1/2}
+h_T^*R{'}_{1/2}^{\dagger})-\sum_{i=0}^l(R{'}_{i+1/2}^\dagger 
R'_{i-1/2}+
h.c.)-\sum_{i=0}^{l-1}M'_{i+1/2}.\end{equation}
We have a longitudinal field at site 1/2, as well as a transverse
field.  Also note that, at the last site, $l+1/2$, there is no field of
either kind.  

\begin{figure}
\epsfxsize=10 cm
\centerline{\epsffile{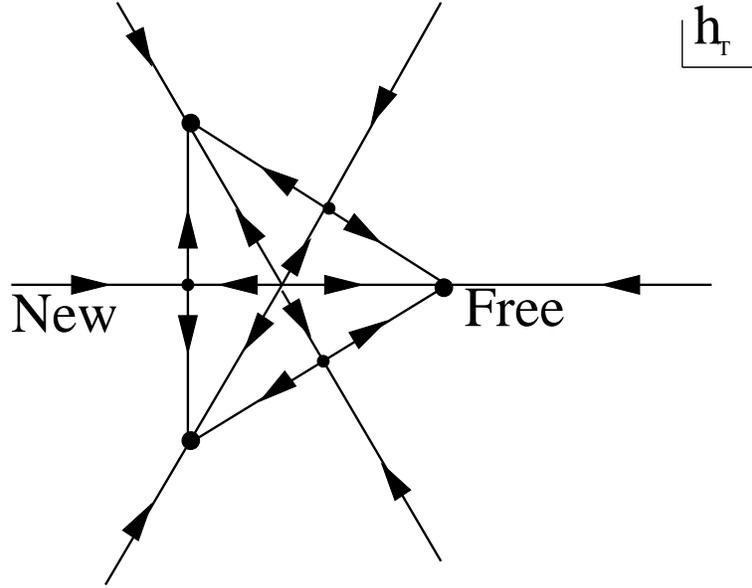}}
\caption{Schematic phase diagram of the quantum chain version of 
the
Potts model with a complex boundary transverse field. Arrows 
indicate direction
of RG flows as the energy scale is decreased.}
\label{fig:RG}
\end{figure}

Consider first the case where $h_T$ is real and positive, for example 
$h_T=1$
corresponding to standard free b.c.s.  The dual model has the 
longitudinal
field term at 1/2:
\begin{equation} -h_T\pmatrix{-1&0&0\cr 0&-1&0\cr 
0&0&2\cr},\end{equation} 
which favours the third
(C) Potts state.  We expect this Hamiltonian to renormalize to the 
fixed (C) b.c..  
The
spin at site $l+1/2$ cannot flip.  We may fix it in the A, B or C state.
This corresponds to a sum of 3 fixed b.c.'s A,B or C.  From the dual
viewpoint the partition function at low energies is
\begin{equation}
Z_{C,A}+Z_{C,B}+Z_{C,C}=\chi_I+\chi_\psi+\chi_{\psi^\dagger}=
Z_{free,free
}
.\end{equation}
This is obviously the correct answer when $h_T=1$ and is a useful 
check on 
duality.  It 
implies that the dual of free is fixed.  Now consider the case where 
$h_T$ is real
and negative.  The dual model has a longitudinal field which favours
states A and B equally.  It should flow to the mixed b.c. AB.  Thus the
partition function is:
\begin{equation} Z_{AB,A}+Z_{AB,B}+Z_{AB,C}=\chi_\epsilon + 
\chi_\sigma 
+
\chi_{\sigma^\dagger}.\end{equation}
  We see from Eq. (\ref{PFs}) that this is $Z_{new,free}$.  This
indicates that we obtain the new b.c. by reversing the sign of the 
transverse
field at the boundary.  We see that the dual of mixed is new. This is 
consistent 
with 
$Z_{new,new}$ in Eq. (\ref{PFs}).
This new b.c. is stable provided that $h_T$ is real and
negative.  There is a discrete symmetry associated with $h_T$ being 
real,
time reversal.  Now let's break this symmetry and give $h_T$ a small
imaginary part, $h_T\to h_T+ih_T'$.  Note that we have not broken 
the $Z_3$
symmetry (in the original formulation).  In the dual picture the
longitudinal field at site 1/2 is:
\begin{equation} \pmatrix{h_T+\sqrt{3}h_T'&0&0\cr 0&h_T-
\sqrt{3}h_T'&0\cr 
0&0&-
2h_T\cr }.\end{equation}
For $h_T>0$ and small $h_T'$ the C state is still favoured.  But for 
$h_T<0$ the
$h_T'$ term breaks the degeneracy between A and B.  We then get a 
flow from
mixed (AB) to fixed (A or B) in the dual picture.  In the original
formulation we get a flow from new to free.  In either picture, the 
flow is driven 
by 
an 
$x=2/5$ boundary operator.  This explains the $Z_3$
symmetric relevant operator at the new fixed point that we were
discussing.  Importantly there is a different symmetry, time reversal, 
which 
forbids
it.  In the complex h-plane the phase diagram can be easily 
constructed.
There are 3 completely stable free fixed points (in the original
formulation) at equal distances from the origin on the positive real 
axis
and at angles $\pm 2\pi /3$.  There are 3 new fixed points at equal
distances from the origin on the negative real axis and at angles $\pm 
\pi
/3$.  These are attractive for flows along rays from the origin but
repulsive for flows perpendicular to these rays.  One can easily 
connect
up these critical points and draw sensible looking flows for the whole
complex plane, as shown in Fig. (\ref{fig:RG}).  Although 3 ``free'' fixed points 
occur in this 
phase 
diagram, they all correspond to the same boundary state.  In fact, arg 
($h_T$) can 
be 
rotated by $2\pi /3$ by a unitary transformation at site 0 by the matrix 
$R_0$.  
Thus the 
3 finite size groundstates (and all excited states) for $h_T$ at the 3 
``free'' fixed 
point 
values, are rigorously identical except for a local change at site 0. The 
spectra, 
with 
any 
given b.c. at $l$ is the same in all 3 cases. Clearly all 3 cases have the 
same long 
distance, low energy properties and should thus be thought of as 
corresponding to 
the 
same fixed point.  Similarly all three ``new'' fixed points are 
equivalent.
It might, in fact, be more appropriate to draw the new fixed point at
$|h_T|=\infty$ rather than at a finite distance from the origin, as in
Fig. (\ref{fig:RG}).  This follows since, in the dual picture, we obtain the mixed
b.c. by eliminating one of the classical Potts states and hence taking
the longitudinal field to $\infty$.  An infinite real negative transverse
field eliminates the symmetric state (1,1,1) at the first site and 
projects
onto the 2 orthogonal states with basis $(1,e^{i2\pi /3},e^{-i2\pi/3})$,
$(1,e^{-i2\pi /3},e^{i2\pi /3})$.

The origin, $h_T=0$, corresponds to
a sum of A,B and C boundary conditions. We may specify a value for 
the Potts 
variable 
at 0 and it is unchanged by the action of the Hamiltonian. The Hilbert 
Space 
breaks 
up 
into 3 sectors depending on which value is chosen. One way of 
checking the 
consistency 
of this is the
duality transformation.  For $h_T=0$ the dual model still has a 
transverse
field at site 1/2 but no longitudinal field.  Thus it corresponds to a
free b.c..  However, in the dual model the b.c. at l+1/2 is a sum of 
A,B,
and C b.c.'s.  Thus we get the same partition function from either 
picture
$$Z_{free,A}+Z_{free,B}+Z_{free,C}.$$

The set of boundary operators at $h_T=0$, is given by the finite size 
spectrum 
with 
a 
sum of A,B and C boundary conditions at each end of the system.  
This gives the 
partition function:
\begin{equation} 
Z=3(Z_{A,A}+Z_{A,B}+Z_{A,C})=3(\chi_I+\chi_\psi+\chi_{\psi^\dagger}).
\label{hT0}\end{equation}
Note that there are 3 zero dimension boundary operators for $h_T=0$.  
One
is the identity.  The other two correspond to a longitudinal field
$h_LR_0+h_L^*R_0^\dagger $.  This should pick out one of the 3 
b.c.'s A,B or C
(for generic phases of $h_L$).  $<R_0>$ takes on a finite value for
infinitesimal $h_L$ corresponding to a 1st order transition. This 
becomes
especially obvious by again using duality but now running the 
argument
backwards.  That is, let's now study the dual model with 0 transverse
field and a small non-zero longitudinal field, $h_L$.  This 
corresponds to the
original model with a transverse field $h_LM_0+$ h.c. but zero 
classical
Potts interaction $R_0^\dagger R_1$ + h.c.  Clearly, $h_L$ produces 
a 1st
order transition in this model since the 1st site is exactly decoupled.
We can diagonalize $M_0$ and 1 or the other of the 3 eigenstates will 
be
the groundstate depending on the phase of $h_L$ (for generic values 
of this
phase).  In the dual model this corresponds to 1st order transitions
between eigenstates of $R_{1/2}$ when a longitudinal field is turned 
on
(with a non-zero classical Potts interaction of order 1).  It is also
clear that there are special values for the phase of $h_L$ for which 2
groundstates remain degenerate so another 1st order transition occurs,
across the negative real $h_L$ axis (and the 2 other axes rotated by 
$\pm 
2\pi/3$.)  

From Eq. (\ref{hT0}), there are 6 relevant boundary operators of 
dimension 2/3 at 
the 
$h_T=0$ fixed point.  We may identify these with these with the 6 
tunneling 
processes 
$A\to B$, $B\to A$, etc.  Imposing $Z_3$ symmetry, no dimension 0 
operators 
and 
only 
2 dimension 2/3 operators are allowed. The latter couple to the 
complex transverse 
field, 
$h_T$.  Thus we see that the flow away from $h_T=0$ to the new or 
free fixed 
points is 
driven by $x=2/3$ operators.

Further insight into the nature of the new fixed point can be gained by 
considering 
again 
the model with no classical Potts interaction between sites 0 and 1, 
$h_T\neq 0$ 
and no 
longitudinal field. Thus we have a Potts chain with a free b.c. at 1 and 
an 
additional 
decoupled Potts spin at 0.
For real positive $h_T$, the decoupled Potts spin has a unique 
symmetric 
groundstate, 
(1,1,1).  In this case, we expect that turning on the classical Potts 
interaction with 
the 1st 
site leads to the free fixed point.  The end spin is simply adsorbed, 
with a flow 
from 
free 
to free.  On the other hand, for real and negative $h_T$, the 
groundstate of the 
decoupled 
spin at 0 is 2-fold degenerate. These 2 states can be chosen to be 
$ (1,e^{2\pi i/3},e^{-2\pi i/3})$ and $(1, e^{-2\pi i/3}, e^{2\pi i/3})$. 
Turning on the classical Potts interaction should now produce a flow 
to the new 
fixed 
point from the above discussion.  Thus we get a flow from a free b.c. 
with a 
decoupled 
system with a 2-fold degeneracy, to the new b.c.  This is somewhat 
like the RG 
flow in 
the S=1/2 Kondo problem, with the 2 states of the decoupled spin in 
the Potts 
model 
corresponding to spin up or down in the Kondo model.  The flow to 
the new fixed 
point 
is analogous to Kondo-screening of the impurity. 
A related problem, an impurity with triangular symmetry coupled to 
conduction 
electrons, was discussed in [\onlinecite{Moustakas}].  The 2-fold 
degeneracy of the 
groundstates 
of the impurity is guaranteed by the $Z_3$ symmetry (for the 
appropriate sign of 
the 
tunneling term) and can lead to 2-channel Kondo behaviour (when 
electron spin 
is 
taken 
into account) without the fine-tuning necessary for ordinary 2-level 
impurities.  
We 
note 
that both these problems correspond to a $Z_3$ symmetric impurity 
coupled to a 
dissipative environment.  In [\onlinecite{Moustakas}] this 
environment is the 
conduction 
electrons;  in our model it is the rest of the Potts chain.

The dual version of this last RG flow is easily constructed. At site 1/2  
there is 
originally a longitudinal
 field but no transverse field.  Thus the system is in a sum of 2 states, 
$A+B$.  
Upon
turning on the transverse field we expect a flow to the mixed state 
$AB$.

\begin{figure}
\epsfxsize=10 cm
\centerline{\epsffile{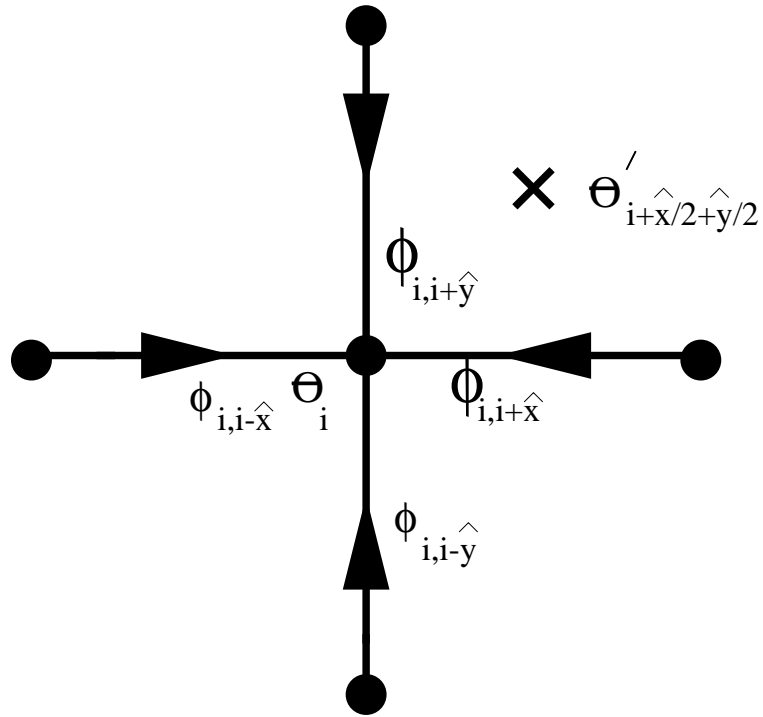}}
\caption{Site, link and dual lattice variables.}
\label{fig:classdual}
\end{figure}

Let us now consider the classical Potts model of Eq. (\ref{class}).  We 
may again 
construct the new boundary fixed point using duality.  The first step is 
to Fourier 
transform the factor associated with each link, $ij$ in the partition 
sum.  Thus we 
introduce a new angular variable, $\phi_{ij}$, taking on values $0$, 
$\pm 2\pi /3$ 
associated with the link ij by:
\begin{equation} e^{J\cos (\theta_i-\theta_j)}=\sum_{\phi_{ij}}
e^{i3\phi_{ij}(\theta_i-\theta_j)/2\pi}Ae^{K\cos(\phi_{ij}),}
\label{FT}\end{equation}
where $A$ is a normalization constant.
We now sum over the original Potts variables, $\theta_i$.  Ignoring 
the 
boundaries, 
for the moment, the sum over the Potts variable at each site gives a 
constraint on 
the 
4 link variables associated with the links terminating at the site.  [See 
Fig. 
(\ref{fig:classdual}).]
\begin{equation} \sum_{\pm}(\phi_{i,i\pm \hat x}+\phi_{i,i\pm \hat 
y})=0\ \ 
(mod \ 
2\pi ),\end{equation} 
where $\phi_{ji}\equiv -\phi_{ij}$.
We may solve these constraints by introducing new angular variables, 
$\theta_i'$ 
(also restricted to the values 0,$\pm 2\pi /3$) on the dual lattice, i.e. 
the centers of 
the squares of the original lattice.    [See Fig. (\ref{fig:classdual})].
Explicitly:
\begin{eqnarray}
\phi_{i,i+\hat y}=\theta_{i+\hat x/2+\hat y/2}'-\theta_{i-\hat x/2+\hat 
y/2}',\nonumber \\
\phi_{i,i+\hat x}=\theta_{i+\hat x/2-\hat y/2}'-\theta_{i+\hat x/2+\hat 
y/2}'.
\end{eqnarray}
The partition function is transformed into:
\begin{equation} Z\propto 
\prod_i\sum_{\theta_i'}e^{\sum_{<i,j>}K\cos 
(\theta_i'-
\theta_j')}.
\end{equation}
Thus we get back the original Potts model with a dual coupling 
constant, $K$.  
The 
critical coupling is given by the self-duality condition, $J=K$, which 
gives:
\begin{equation} J_c={2\over 3}\ln (1+\sqrt{3}).\end{equation}

\begin{figure}
\epsfxsize=10 cm
\centerline{\epsffile{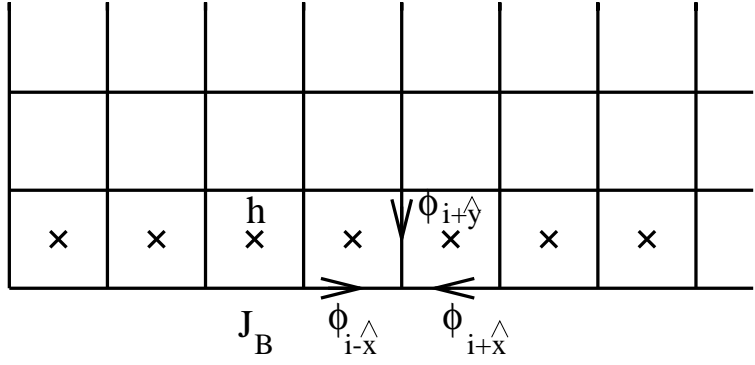}}
\caption{Boundary variables.}
\label{fig:dualbound}
\end{figure}

Now consider the system with a free boundary along the x-axis with a 
boundary 
Potts 
interaction $J_B$ (and no fields at the boundary).  Consider summing 
over the 
Potts 
variable $\theta_i$ at the boundary, as indicated in Fig. 
(\ref{fig:dualbound}).
This gives the constraint:
\begin{equation}
\phi_{i,i+\hat y}+\phi_{i,i+\hat x}+\phi_{i,i-\hat x}=0.\end{equation}
Writing $\phi_{i,i+\hat y}$ in terms of the dual variables, this 
becomes:
\begin{equation} \theta_{i-\hat x/2+\hat y/2}'-\theta_{i+\hat x/2+\hat 
y/2}'+ 
\phi_{i,i+\hat x}+\phi_{i,i-\hat x}=0.\end{equation}
We may solve this equation for all sites, i, along the boundary by:
\begin{equation} \phi_{i,i+\hat x}=\theta_{i+\hat x/2+\hat 
y/2}'.\end{equation}
Thus the edge of the dual lattice is at $y=1/2$.  In addition to the
bulk Potts interaction of strength $K$, given by
Eq. (\ref{FT}), there is an additional classical boundary term in the 
dual 
Hamiltonian:
\begin{equation} -\beta H_{\hbox{field}}=h\sum_j\cos 
\theta_{j,0}',\end{equation}
with the dual boundary field, h, determined by the boundary 
interaction:
\begin{equation} e^{J_B\cos (\theta_i-\theta_{i+\hat 
x})}=\sum_{\phi_{i,i+\hat 
x}}
e^{i3\phi_{i,i+\hat x}(\theta_i-\theta_{i+\hat 
x)}/2\pi}Ce^{h\cos(\phi_{i,i+\hat 
x})},
\end{equation}
for some constant, $C$.
This gives the condition:
\begin{equation} e^{3J_B/2}={e^h+2e^{-h/2}\over e^h-e^{-
h/2}}.\end{equation}
This equation has the annoying feature that, for real $J_B$ and $h$, 
there are only 
solutions for $h$ and $J_B>0$, with $h$ running from $\infty$ to $0$ 
as $J_B$ 
runs 
from $0$ to $\infty$.  From our analysis of the quantum model we 
expect that the 
new critical point occurs when the dual model has a real negative $h$.  
This 
requires 
a complex $J_B$, Im($J_B)=2\pi /3$.
Noting that the ratio of Boltmann weights for $\theta_i-\theta_{i+\hat 
x}=0$ or 
$\pm 
2\pi /3$ is $e^{3J_B/2}$, we see that this implies  real but negative 
Boltzmann 
weights. In particular, we way regard the new fixed point as corresponding 
to an infinite 
negative $h$; this corresponds to 
\begin{equation} e^{3J_B/2}=-2.\end{equation}
In the quantum model, discussed above, this limit eliminates the 
symmetric state
(1,1,1) on the first site, projecting onto the 2 orthogonal states.
The same projection is realized in the standard transfer matrix 
formalism
for the classical Potts model.
The Potts model with negative Boltzmann weights in the bulk occurs 
quite 
naturally in 
the cluster
formulation based on the high temperatures expansion.\cite{Saleur2}

We note that the values of the ``groundstate degeneracies'' of the 
various fixed
points, $<\alpha |0,0>$ are given by:
\begin{equation} g_A=N,\ \ g_{AB}=N\lambda^2,\ \ 
g_{free}=N\sqrt{3},\ \ 
g_{new}=N\sqrt{3}\lambda^2 .\end{equation}
Noting that 
\begin{equation} 1<\lambda^2={1+\sqrt{5}\over 
2}<\sqrt{3},\end{equation}
we see that:
\begin{equation}
g_A<g_{AB}<g_{free}<g_{new}<3g_A<2g_{free}.\end{equation}
Thus all RG flows that we have discussed are consistent with the ``g-
theorem''\cite{Affleck4}
(or ``g-conjecture'' as it is more accurately referred to).  g always 
decreases under 
an RG flow.  We also note that the various flows which are related by 
duality have 
the same 
ratios of g-factors:
\begin{eqnarray} {g_{new}\over g_{free}}&=&{g_{AB}\over 
g_A}=\lambda^2\nonumber \\
{3g_A\over g_{free}}&=&{g_{free}\over g_A}=\sqrt{3} \nonumber 
\\
{3g_A \over g_{new}}&=&{g_{free}\over g_{AB}}={\sqrt{3}\over 
\lambda^2}
\nonumber \\
{2g_{free}\over g_{new}}&=&{2g_A\over g_{AB}}={2\over 
\lambda^2}.
\end{eqnarray}

This research was begun while all three authors were visitors at
the Institute for Theoretical Physics, Santa Barbara.
The research of IA and MO was supported by NSERC of Canada and 
the
Killam Foundation.  That of HS by the DOE, the NSF and the Packard
Foundation.
\appendix \section{$\lowercase{g}$-theorem for fusion}
At present, the most general and systematic method to construct a new
boundary state is fusion. For rational CFTs
(with a finite number of conformal towers), fusion is quite
a powerful method.
Empirically it has been recognized that fusion is a kind of
irreversible process: when a boundary state $B$ is obtained by
 fusion from another boundary state $A$, fusion on $B$ does
not generally gives $A$. (Sometimes it does.)
This irreversibility reminds us of the ``$g$-theorem'' which states
that the groundstate degeneracy $g$ of the system always decreases
along the boundary renormalization group flow.\cite{Affleck4}
Actually, here we prove that the irreversibility of fusion is
also related to the groundstate degeneracy $g$.
Amusingly, the ``direction'' is opposite to that of
renormalization group flow. We state the following:

\begin{description}
\item[Theorem]
We consider a unitary rational CFT.
Let $B$ be a boundary state obtained by fusion from the boundary
state $A$. The groundstate degeneracy of $B$ is always greater
than or equal to that of $A$.
\end{description}

To prove the theorem, first let give us the definition of the
groundstate degeneracy. Given a boundary state $| X \rangle$,
the groundstate degeneracy of the state $g_X$ is given by
the following:
\begin{equation}
  g_X = \langle 0 | X \rangle,
\label{degdef}\end{equation}
where $| 0 \rangle$ is the groundstate of the system and
we choose the overall phase of $| X \rangle$ so that $g_X$ is
positive.
For unitary CFTs, the groundstate corresponds to the identity
operator with conformal weight $0$. We denote this identity primary
as $0$.  The definition of Eq. (\ref{degdef}) follows from the fact that 
the partition
function is proportional to this matrix element in the limit of an
infinite length system.

On the other hand, a general relation for
fusion\cite{Cardy,Affleck4}
reads:
\begin{equation}
  \langle  a | B \rangle = \langle a | A \rangle \frac{S^a_c}{S^a_0},
\end{equation}
where $a$ represents an aribitrary primary field, $c$ is the primary
used for fusion from $A$ to $B$, and $S^x_y$ is the modular $S$-
matrix
element for primaries $x$ and $y$.
The special case $a=0$ (identity) gives the relation between
the degeneracies:
\begin{equation}
  \frac{g_B}{g_A} = \frac{S^0_c}{S^0_0} .
\label{eq:gBgA1}
\end{equation}

Now we employ the Verlinde formula\cite{Verlinde,Cardy}:
\begin{equation}
  \sum_b S^a_b N^b_{cd} = \frac{S^a_c S^a_d}{S^a_0}.
\end{equation}
The special case $a=0$ and $d=c'$ ($c'$ is the conjugate of $c$),
combined with eq.~(\ref{eq:gBgA1}) gives
\begin{eqnarray}
  \left| \frac{g_B}{g_A}\right|^2 &=& \frac{1}{S^0_0} \sum_b S^0_b 
N^b_{c'c}
\nonumber \\
&=& 1 + \sum_{b \neq 0} \frac{S^0_b}{S^0_0} N^b_{c'c},
\end{eqnarray}
where we used the fact the operator product expansion between $c$
and its conjugate $c'$ always contains the identity operator.

Since the fusion rule coefficients $N^b_{cc}$ are nonnegative 
integers,
the theorem follows if $S^0_b/ S^0_0>0$.
Actually, it is known that $S^0_b > 0$ for any primary $b$, 
proved as follows.\cite{Difrancesco}
Consider the character $\chi_b(\tilde q)$.
By modular transformation,
\begin{equation}
  \chi_b(\tilde q) = \sum_e S^e_b \chi_e(q) .
\end{equation}
When evaluating the limit $q \to 0$,
the right-hand side is dominated by the lowest power of
$q$.
Thus $\chi_b(\tilde q) \sim S^0_b q^{-c/24}$. (Here $c$ in the 
exponent
is the central charge of the CFT.)
Since the left-hand side and $q^{-c/24}$ are both positive, $S^0_b > 
0$.
Thus the theorem is proved.

Our theorem is of course consistent with all known cases, including
boundary states of the Ising and Potts models.
When there is an RG flow between two boundary states, our theorem
implies that the direction of the fusion rule construction is opposite.
Namely, we can obtain an unstable boundary state from a more stable
boundary state, but the reverse is not possible.
However, there can be an exception: if there are some extra degrees
of freedom, the RG flow can be in the same direction as the fusion.
An example of this is the Kondo effect; the screened state, which
has larger groundstate degeneracy than the original state, is
constructed by fusion. However, if we take the degeneracy due to
the impurity spin into account, the total degeneracy is smaller
in the screened state. Thus the RG flow occurs from the unscreened
to screened state.

That fusion generates rather opposite ``flow'' to the RG one
makes it somewhat difficult to understand the physical meaning of
 fusion, which is a more or less abstract mathematical manipulation.
Perhaps the best intuition is gained again from the example of the 
Kondo
effect. Namely, fusion roughly corresponds to an absorption of
some degree of freedom by the boundary.
Considering the generality of the present result, it is tempting
to imagine some deeper connection with the ``$g$-theorem''
on the RG flow.

\begin{table} \caption{The modular S-matrix for Virasoro characters 
(multiplied 
by 
$2/N^2$).  Characters are labeled by their Kac labels (n,m) (and by 
their heighest 
weight).} \label{tab:SM}
\begin{tabular}{|r||r|r|r|r|r|r|r|r|r|r|} &$11(0)$ & $41(3)$ & $21(2/5)$ & 
31(7/5)& 
$43(2/3)$ 
&33(1/15)&44(1/8)&42(13/8)&22(1/40)&32(21/40)\\
\tableline 
$11(0)$&$1$&$1$&$\lambda^2$&$\lambda^2$&$2$&$2\lambda^2
$&
$\sqrt{3}$&$\sqrt{3}$&$\sqrt{3}\lambda^2$&$\sqrt{3}\lambda^2$\\
$41(3)$&$1$&$1$&$\lambda^2$&$\lambda^2$&$2$&$2\lambda^2
$&
$-\sqrt{3}$&$-\sqrt{3}$&$-\sqrt{3}\lambda^2$&$-
\sqrt{3}\lambda^2$\\
$21(2/5)$&$\lambda^2$&$\lambda^2$&$-1$&$-
1$&$2\lambda^2$&$-2$&
$-\sqrt{3}\lambda^2$&$-
\sqrt{3}\lambda^2$&$\sqrt{3}$&$\sqrt{3}$\\
$31(7/5)$&$\lambda^2$&$\lambda^2$&$-1$&$-
1$&$2\lambda^2$&$-2$&
$\sqrt{3}\lambda^2$&$\sqrt{3}\lambda^2$&$-\sqrt{3}$&$-
\sqrt{3}$\\
$43(2/3)$&$2$&$2$&$2\lambda^2$&$2\lambda^2$&$-2$&$-
2\lambda^2$&
$0$&$0$&$0$&$0$\\
$33(1/15)$&$2\lambda^2$&$2\lambda^2$&$-2$&$-2$&$-
2\lambda^2$&$2$&
$0$&$0$&$0$&$0$\\
$44(1/8)$&$\sqrt{3}$&$-\sqrt{3}$&$-\sqrt{3}\lambda^2$ 
&$\sqrt{3}\lambda^2$&$0$&$0$&$-\sqrt{3}$&$\sqrt{3}$
&$\sqrt{3}\lambda^2$&$-\sqrt{3}\lambda^2$\\
$42(13/8)$&$\sqrt{3}$&$-\sqrt{3}$&$-\sqrt{3}\lambda^2$ 
&$\sqrt{3}\lambda^2$&$0$&$0$&$\sqrt{3}$&$-\sqrt{3}$
&$-\sqrt{3}\lambda^2$&$\sqrt{3}\lambda^2$\\
$22(1/40)$&$\sqrt{3}\lambda^2$&$-\sqrt{3}\lambda^2$&$\sqrt{3}$ 
&$-
\sqrt{3}$&$0$&$0$&$\sqrt{3}\lambda^2$&$-\sqrt{3}\lambda^2$
&$\sqrt{3}$&$-\sqrt{3}$\\
$32(21/40)$&$\sqrt{3}\lambda^2$&$-
\sqrt{3}\lambda^2$&$\sqrt{3}$ &$-
\sqrt{3}$&$0$&$0$&$-\sqrt{3}\lambda^2$&$\sqrt{3}\lambda^2$
&$-\sqrt{3}$&$\sqrt{3}$\\
\end{tabular} \end{table}
\begin{table}
\caption{Fusion rules for extended operator algebra. Fusion rules not 
shown are
the standard ones for the Potts model [\protect\onlinecite{Cardy}].}
\label{tab:fus}
\begin{tabular}{|c||c|c|c|c|}
&$I$ or $\psi$ or $\psi^\dagger$&$\epsilon$ or $\sigma$
or $\sigma^\dagger$&${\cal O}_{44}+{\cal O}_{42}$&
${\cal O}_{22}+{\cal O}_{32}$\\
\tableline
${\cal O}_{44}$ or ${\cal O}_{42}$&${\cal O}_{44}+{\cal 
O}_{42}$
&${\cal O}_{22}+{\cal O}_{32}$&$I + \psi + 
\psi^\dagger$&$\epsilon + \sigma 
+
\sigma^\dagger$ \\
${\cal O}_{22}$ or ${\cal O}_{32}$&${\cal O}_{22}+{\cal 
O}_{32}$&${\cal 
O}_{44}+{\cal O}_{42}+{\cal O}_{22}+{\cal O}_{32}$&$\epsilon 
+ \sigma 
+\sigma^\dagger $&$I + \psi
+\psi^\dagger +\epsilon +\sigma + \sigma^\dagger$ \\
\end{tabular}
\end{table}

\end{document}